\documentclass[aps, pre, twocolumn,a4paper,showkeys,showpacs]{revtex4-1}
\usepackage{graphicx}
\usepackage{amsmath}
\usepackage{bm}
\usepackage{color}
\usepackage{natbib}
\usepackage{comment}

\begin{document}
\title{Emergence of collective propulsion through cell-cell adhesion}
 
\author{Katsuyoshi Matsushita}

\affiliation{Department of Biological Sciences, Osaka University, Toyonaka, Osaka, Japan}

\begin{abstract}
The mechanisms driving the collective movement of cells remain poorly understood. To contribute toward resolving this mystery, a model was formulated to theoretically explore the possible functions of polarized cell-cell adhesion in collective 
cell migration. The model consists of an amoeba cell with polarized cell-cell adhesion, which
is controlled by positive feedback with cell motion. This model cell has no 
persistent propulsion, and therefore exhibits a simple random walk when in isolation. However, at high density, these cells acquire collective propulsion 
and form ordered movement. This result suggests that cell-cell
adhesion has a potential function, which induces collective propulsion with persistence.
\end{abstract}

\maketitle

Collective cell migration is an indispensable element for various developmental, physiological, and 
pathological processes \cite{Weijer:2009, Friedl:2009, Rorth:2009}. However, the guiding mechanisms driving the movement of cells during migration are not sufficiently 
understood. Various biological hypotheses have been
proposed to elucidate these mechanisms \cite{Haeger:2015}, which have been examined in the field of physics \cite{Sherratt:1990, 
Szabo:2006, Lee:2011a, Lee:2011b, Vicsek:2012, Basan:2013, Marchetti:2013, 
Sepulveda:2013, Li:2014, Londono:2014, Camley:2016}. One of the most widely investigated hypotheses
is based on the concept of a leader cell that differentiates to lead other cells \cite{Kabla:2012}. 
Another major hypothesis is extracellular matrix (ECM) leading, including durotaxis 
\cite {Lo:2000} and haptotaxis \cite{Carter:1967}. Along with these models, various other hypothetical 
guiding mechanisms can qualitatively reproduce many aspects of collective cell migration.

Among these mechanisms, the most simple guiding principle is one in which homogeneous cells 
mutually lead themselves independently of the ECM, which is referred to here as the ``mutual leading mechanism.'' In spite of the simplicity of this type of guiding, it induces rich 
collective behavior \cite{Mehes:2013}.
In these behaviors, leading is based on cell-cell communication. Chemotaxis is a major 
communication tool used for cellular interactions \cite{Pfeffer:1884}
as observed in the aggregation of {\it Dictyostelium discoideum} 
\cite{Bonner:2009} and in contact inhibition of the locomotion of 
neural crest cells \cite{ Carmona-Fontaine:2008, Camley:2016}. 
Therefore, investigations of the mutual leading mechanism conducted to date 
have mainly focused on the chemotactic response of cells \cite{Swaney:2010}.

\begin{figure}[t] 
\begin{center} 
\includegraphics[width=1.0\linewidth]{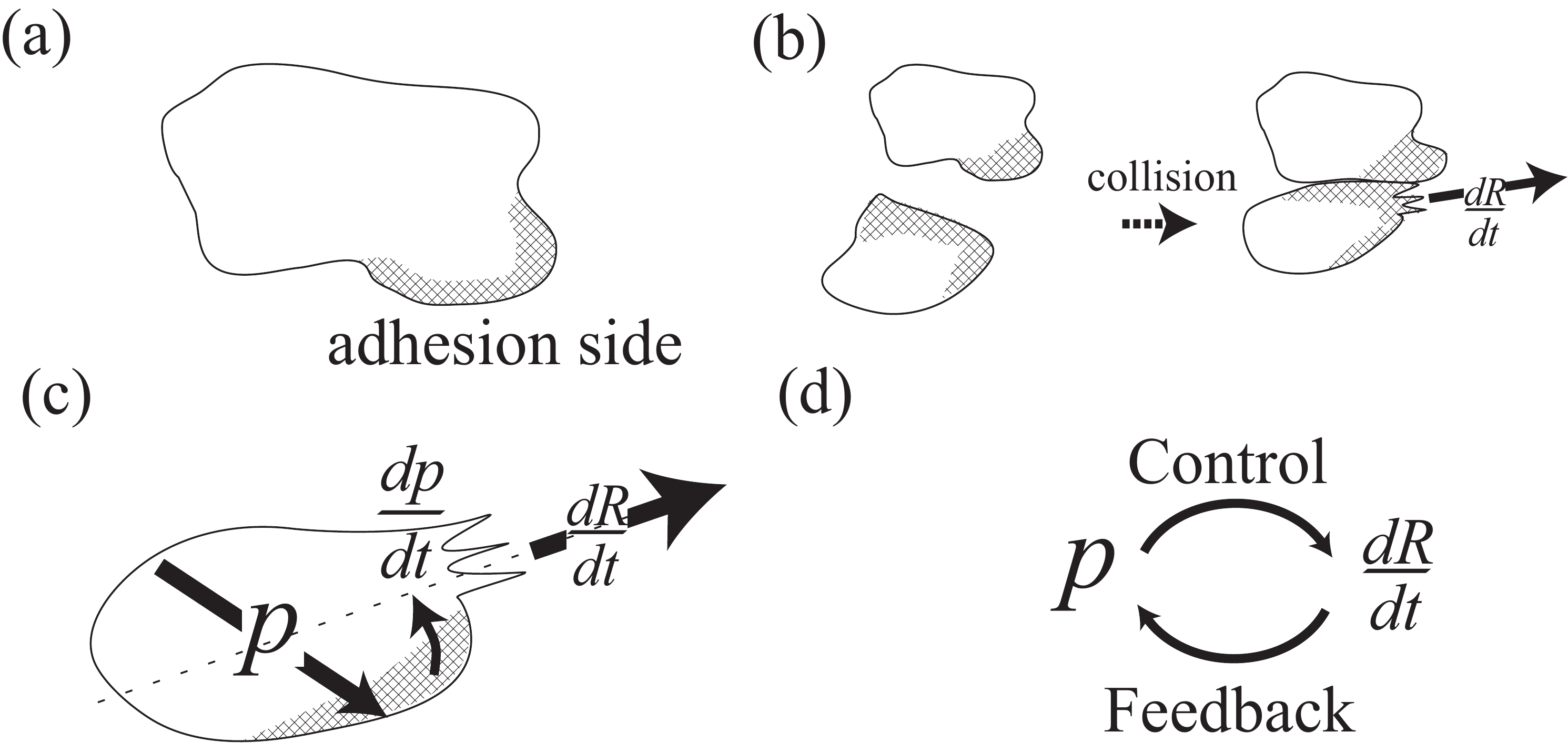} 
\end{center}
\caption{
(a) Schematic view of a model cell with polarized cell-cell adhesion. 
The shaded region 
represents a high-strength region of cell-cell adhesion. 
(b)
The collision process of two cells. The arrows represent cell movement.
$d{\bm R}/{dt}$ represents the direction of cell motion. 
The saw tooth shape of the bottom cell represents the protrusion that induces
cell movement.
The two cells collide
and are then bound through cell-cell adhesion. As a result, the cells move in essentially the same direction.
(c) 
Dynamics of polarized cell-cell adhesion. 
The arrow of ${\bm p}$ represents the direction of 
single-side polarized cell-cell adhesion and 
the curved arrow represents $d{\bm p}/{dt}$ according to 
Eq.~\ref{eq:eq_polarity}. The dotted line indicates the direction of $d{\bm R}/{dt}$.
(d) Schematic relation between ${\bm p}$ and $d{\bm R}/{dt}$.
 }\label{fig:model}
\end{figure}

Another possible communication tool is cell-cell adhesion 
\cite{Takeichi:2014}. In contrast to the in-depth understanding of the functions
of chemotaxis in collective cell migration, knowledge of the role of cell-cell adhesion 
is limited. In particular, although the role of cell-cell adhesion in the leader cell mechanism has been recently clarified 
\cite{Kabla:2012, Dumortier:2012, Cai:2014}, its role in the mutual leading mechanism remains largely unclear.
Cell-cell adhesion can possibly act as a driving force for
collective behavior \cite{Friedl:2003}, including the alignments of {\it Dictyostelium 
discoideum} \cite{Beug:1973, Muller:1978} and
the neural crest \cite{Theveneau:2010}. 
To intuitively consider the functions of cell-cell adhesion in these types of cells, I begin with a thought experiment
using a model amoeba cell population that exhibits cell-cell adhesion, as shown in Fig.~\ref{fig:model}(a). 
When a cell leads other cells to align their directions of movement via cell-cell adhesion, 
single-side polarization in cell-cell adhesion
is necessary. This is because cells cannot
indicate a certain direction of movement through isotropic cell-cell adhesion. 
This type of polarized cell-cell adhesion promotes the protrusion of other cells 
toward the direction of adhesive polarization in collision processes, 
as shown in Fig.~\ref{fig:model}(b).
The directed protrusion then synchronizes the movement of gathered cells, and is ultimately expected to
bring about collective migration. 

To test this expectation, the potential of polarized cell-cell adhesion as a 
communication tool in collective cell migration is theoretically explored.  
Indeed, such polarization does appear to result in the gathering of model cells while further providing them with collective propulsion 
with persistence, even when a cell in isolation only exhibits random-walk movement without persistence. 
Through this propulsion, the model cells switch their motion from random
to collective with a sufficient strength of polarized cell-cell adhesion.

Let us first consider the case of cells with the unit vector of
polarity direction ${\bm p}$ in cell-cell adhesion. To properly lead other cells, 
a cell should adjust ${\bm p}$ using the 
information of movement. A hypothesis for this adjustment mechanism is that  
$\bm p$ simply follows the protrusion or the inducing motion of a cell, as shown in 
Fig.~\ref{fig:model}(c). This is formulated by 
\begin{eqnarray}
\frac{d {\bm p}}{d t} 
= - \eta {\bm p} \times \left( {\bm p} \times \frac{{d \bm R}}{dt} \right)
\label{eq:eq_polarity}
\end{eqnarray} 
Here, $t$ is time and $\bm R$ is the position of the cell. This equation
induces a high correlation between $\bm p$ and $d{\bm R}/dt$ through a 
positive feedback loop (Fig.~\ref{fig:model}(d)). This 
hypothetical feedback has been observed during the early development of {\it Dictyostelium 
discoideum} 
\cite{Sesaki:1996}. Eq.~\eqref{eq:eq_polarity} is employed as a working conjecture for mechanism exploration.

To facilitate the theoretical exploration, an artificial model of cultured 
cells on a two-dimensional (2D) medium is considered, according to the 2D cellular Potts 
model \cite{Graner:1992, Anderson:2007}.  This model generates a probable amoeba
cell configuration by a Monte Carlo method and enables the sampling of
probable cell configurations. In this model, the cell configurations are represented 
by Potts states $m({\bm r})$, representing the state
at a site ${\bm r}$ on a square lattice with a linear dimension of $L$.
The set of all $m({\bm r})$ values is denoted by $\{m({\bm r})\}$.
 $m({\bm r})$ takes on a number in $\{0, 1, \dots, N\}$. When 
$m(\bm r)$ = 0,  $\bm r$ is empty; otherwise, $\bm r$ is occupied by
the $m(\bm r)$th cell. Hence, the domain of $m(\bm r)$ = $m$ determines the 
shape of the $m$th cell.  $N$ is the number of 
cells. For simplicity in the present exploration, a constant $N$ is assumed by ignoring 
the effects of cell division and death. 

Using this model, the possible configurations of cells are sampled
based on Monte Carlo simulation with a probability of realization for $\{m({\bm r})\}$. The probability 
$P(\{m({\bm r}) \})$ is proportional to $\exp(-\beta {\cal H}(\{m({\bm r}) \}))$. Here, 
$\beta$ is a parameter of cell motility and ${\cal H}(\{m({\bm r})\})$ is energy 
defined by
\begin{eqnarray}
{\cal H}(\{m({\bm r})\}) =\sum_{\left<\bm r, \bm r' \right>} 
J_{\bm r\bm r'}\left[1-\delta_{m(\bm r)m(\bm r')}\right]
\nonumber \\ + \kappa \sum_{m=1}^{N_{\rm Cell}} 
\left(V_m - V\right)^2.\label{eq:Hamiltonian}
\end{eqnarray}
The first term on the right-hand side represents energies derived from the tension of 
the cell periphery in the medium and the tension of cell-cell contact 
\cite{Glazier:1993, Graner:1993}. In this term, the summation of 
$\left<\bm r, \bm r' \right>$ is taken over all neighboring site pairs, consisting of the nearest and next-nearest 
neighbor site pairs 
\cite{Graner:1992}. $\delta_{mm'}$ is the Kronecker delta. 
$J_{{\bm r}{\bm r'}}$ is the strength of the interface tension between
$\bm r$ and $\bm r'$.
The second term on the right-hand side represents the area stiffness energy. 
By this term, the area of the $m$th cell, $V_m$ $=$ $\sum_{\bm r} \delta_{mm({\bm r})}$, 
is maintained to be a certain value, $V$. Here, $\kappa$ is the stiffness of the area. 

This cellular Potts model has been used for expressing various polarized cell-cell adhesion events
\cite{Zajac:2002, Vroomans:2015, Matsushita:2017}. To express the single-side polarized 
cell-cell adhesion shown in Fig.~\ref{fig:model}(a), as $J_{{\bm r}{\bm r'}}$ in 
Eq.~\eqref{eq:Hamiltonian}, Eq.~(12) from
Ref.~\cite{Matsushita:2017} is adopted:
\begin{eqnarray}
J_{{\bm r}{\bm r'}} 
= J_{m(\bm r)m(\bm r')} - J_{\rm p}w_m(\bm r)w_{m'}(\bm r'), 
\label{eq:renormalized-coupling}
\end{eqnarray}
where $w_m(\bm r)$ = $(1+{\bm p}_m \cdot {\bm e}_{m}({\bm r}))$
$\eta_{m0}$$/2$, $J_{mm'}$ = $J_{\rm CM}$ $[\delta_{m0}\eta_{m'0}$
$+$ $\eta_{m0}\delta_{m'0}]$ $+$ $J_{\rm CC}$ $\eta_{m0}$
$\eta_{m'0}$, and $\eta_{ab}$ = $1 - \delta_{ab}$. $J_{\rm CM}$ is
the tension of the cell periphery, $J_{\rm CC}$ is the strength of isotropic
cell-cell adhesion, and $J_p$ is the strength of polarized cell-cell 
adhesion. 
 ${\bm p}_m$ is a unit vector representing polarization of the $m$th cell in cell-cell
adhesion. ${\bm e}_m(\bm r)$ is a unit vector from the center of the 
$m$th cell, ${{\bm R}_m}$, to a position on the cell periphery, ${\bm r}$. 
Concretely, it is defined by $({\bm r}$ $-$ ${{\bm R}_m})$ $/$ 
$|{\bm r}$ $-$ ${{\bm R}_m}|$. Here, ${\bm R}_m$ $=$ 
$\sum_{{\bm r}\in \Omega_m}{\bm r}$ $/$ $V_m$, where $\Omega_m$
is the set of all the sites occupied by the $m$th cell. 
In Eq.~\eqref{eq:renormalized-coupling}, 
${\bm p}_m$ obeys the 2D version of Eq.~\eqref{eq:eq_polarity}:
\begin{eqnarray}
 \frac{d{\bm p}_m}{dt} 
= \eta\left[\frac{d{\bm R}_m}{dt}
-\left(\frac{d{\bm R}_m}{dt}\cdot{\bm p}_m\right){\bm p}_m\right]. 
\label{eq:eq-m-mu}
\end{eqnarray}

With this model,
the time series of cell configuration is generated by the following conventional
Monte Carlo process. In this process,
a Monte Carlo step is iterated and produces amoeba cell motion.
The single Monte Carlo step consists of 16 $\times$ $L^2$ copies of a state
from a source site ${\bm r}'$ to a trial site ${\bm r}$. 
For each copy, this trial site $\bm r$ is randomly chosen among all sites. 
Then, a source site, ${\bm r}'$, is 
selected randomly among neighboring sites of ${\bm r}$. 
The state copy of $m({\bm r}')$ from ${\bm r}'$ to $\bm r$
 is accepted with the Metropolis probability of
\begin{align}
&P(\{m_{\rm c}(\bm r)\}|\{m({\bm r}) \}) \nonumber 
\\ &= \min\left[1, P(\{m_{\rm c}(\bm r)\})/P(\{m(\bm r)\})\right]. 
\end{align}
Otherwise, it is rejected. Here, $\{m_{\rm c}(\bm r)\}$ is the state in which
the state is copied from $\bm r'$ to $\bm r$.

For the integration of Eq.~\eqref{eq:eq-m-mu}, the
Euler method is employed. In addition, the adiabatic approximation is employed, where
the equation is integrated only between two consecutive Monte
Carlo steps. This approximation is based on the assumption that
the change of ${\bm p}_m$ is much slower than the rate of a single flip. 
To maintain consistency between time scales in Eq.~\eqref{eq:eq-m-mu},
it is assumed that ${\bm R}_m$ is an adiabatic value and is a 
constant during each Monte Carlo step. ${\bm R}_m$ is also 
calculated with each integration of Eq.~\eqref{eq:eq-m-mu}.

Note that this model does not include the propulsion term
of individual cells. Namely, these model cells only exhibit random-walk movement
without persistence when in isolation. Nevertheless, at high density, these cells may collectively acquire 
propulsion with persistence if the polarized cell-cell 
adhesion functions as a leading communication tool.  

Even if cells acquire the collective propulsion with persistence, 
it is only expected for a limited set of model parameters. In particular,
the area fraction of cells $\phi$ $=$ $NV/L^2$ is a control factor
because a small $\phi$ reduces the cell-cell adhesion effects. Therefore, 
the appropriate value of $\phi$ is first determined for the present exploration. 
As a possible probe of this propulsion, the average value 
of $\bm p_m$ is considered, 
\begin{eqnarray}
P = \frac{1}{TN}\left|  \sum_{m} \sum_{t=t_0}^{t_0+T} {\bm p}_m(t) \right|.
\end{eqnarray}
Since $\bm p_m$ reflects cell motion through Eq.~\eqref{eq:eq-m-mu},
$P$ is expected to reflect the emergence of collective propulsion. 
Here, $t_0$ is a starting time of the time average of ${\bm p}_m$.  To access a steady state, 
we can set $t_0$ = 5 $\times$ $10^5$ steps,  
and simulate the relaxation from cells forming a single 
aggregation with a random configuration of ${\bm p}_m$s up to the time.
We can also set $T$ = 2 $\times$ $10^5$ to calculate the mean 
square displacement, $D^2$, over the long term, as described below. Here, the proper value of $\phi$ is explored by 
calculating the $\phi$-dependence of $P$ in the corresponding range of $N$ from 1 
to 512 with $V$ = 64 and $L$ = 196. In this case, the periodic boundary condition is adopted,
which enables cells to freely move through the boundary. 

In this simulation, the adhesion parameters are set to $J_{\rm CM}$ = 2.0, $J_{\rm CC}$ = 
5.0, and $J_p$ = 2.0 to represent the cell model shown in Fig.~\ref{fig:model}(a). With these parameters, 
cells extend their interface during contact between their front sides and contract their interface during contact of their rear sides. 
Here, the front side of a cell is defined according to the peripheral edge of the cell in the direction of polarized cell-cell adhesion; the rear side is that opposite to the front side. 
For numerical stability, $\eta$ = 0.1, and $\kappa$ = 1 and $\beta$ = 0.2 or 
 $\beta$ = 0.5 are chosen. 

\begin{figure}[t] 
\begin{center} 
\includegraphics[width=1.0\linewidth]{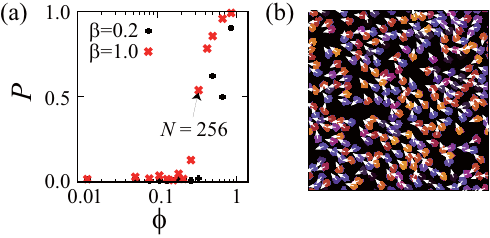} 
\end{center}
\caption{
(a) Order parameter of polarity $P$ as a function of the area
fraction $\phi$.  (b) Snapshot of $\{m(\bm r)\}$ and $\bm p_m$s
at $N$ = 256 ($\phi$ = 42\%) and $\beta$ = 0.5. 
The colored region indicates cells. Different colors represent different cells.
The black region represents empty space. White arrows represent the direction
of polarized cell-cell adhesion.}\label{fig:order_of_polarity}
\end{figure}

$P$ is plotted as a function of $\phi$ in Fig.~\ref{fig:order_of_polarity}(a). For small $\phi$, $P$
takes on a small value. As $\phi$ increases up to around 0.3, $P$ rapidly increases. With further 
increases in $\phi$, $P$ gradually reaches unity, indicating the progression of $\bm p_m$ ordering. This 
transition of $P$ reflects the collective motion occurring for large $\phi$ and its underlying propulsion. 
To gain insight into $\bm p_m$s, a snapshot of $\{m(\bm r)\}$ and $\bm p_m$s is shown in a relaxed state for $N$ = 256
($\phi$ = 42\%) in Fig.~\ref{fig:order_of_polarity}(b). The polarities of ${\bm p}_m$s are indicated by 
arrows that exhibit ordering.  

Next, to address the propulsion of this collective motion and its persistence, the $D^2$ is calculated and averaged across the cells. 
\begin{eqnarray}
D^2 = \frac{1}{N}\sum_m \left|{\bm R}_m(t_0 + t) - {\bm R}_m( t_0) \right|^2
\end{eqnarray}
When cells have propulsion with persistence, they exhibit ballistic motion during a short period; therefore, $D^2$ behaves as $D^2$ $\sim$ $t^2$. Otherwise, the cells diffusively
move, and therefore $D^2$ behaves as $D^2$ $\sim$ $t$. 
Here, we will concentrate on the case of $\beta$ = 0.5
to clearly observe the stable motion 
in comparison with the case of low $\beta$.

\begin{figure}[t] 
\begin{center} 
\includegraphics[width=1.0\linewidth]{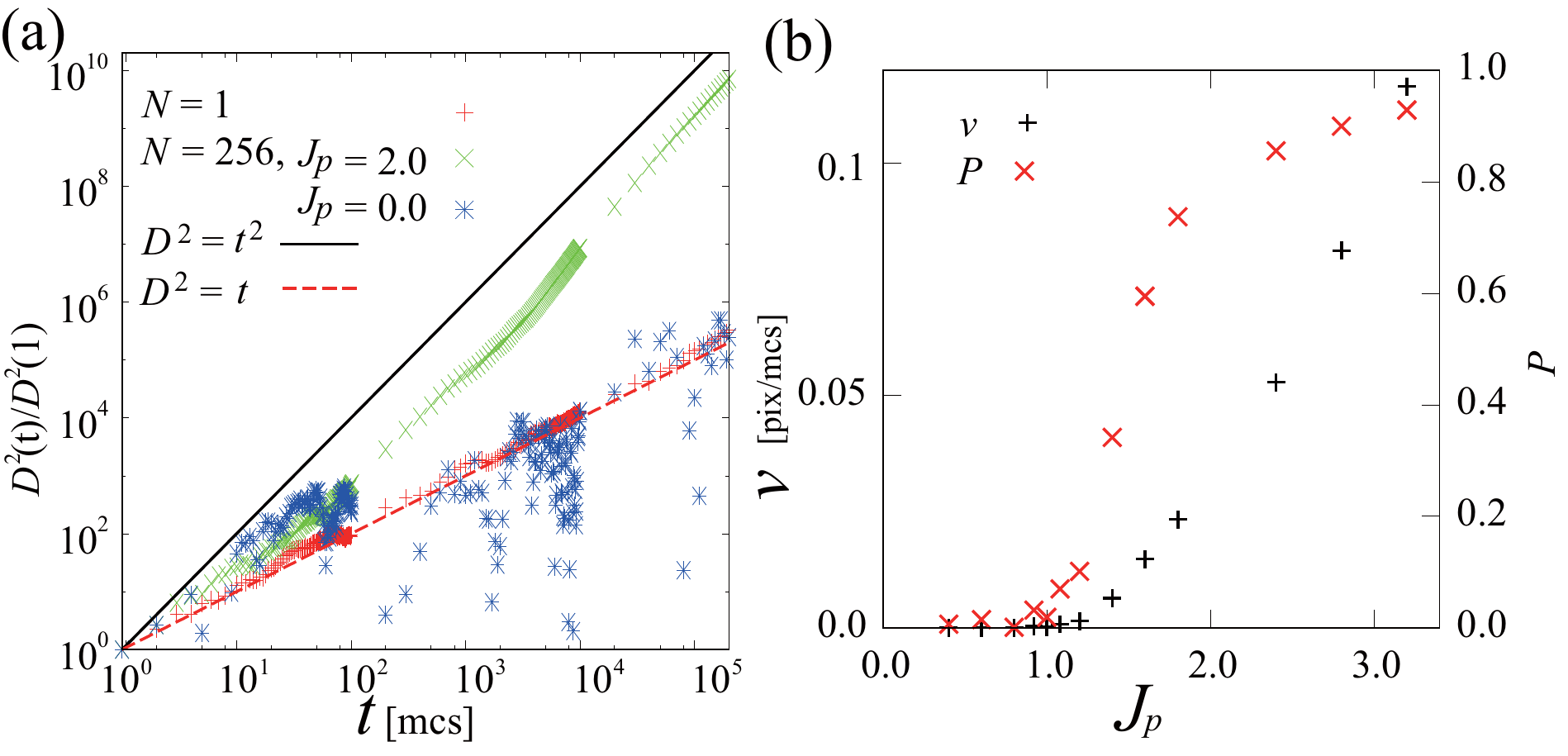} 
\end{center}
\caption{
(a) $D^2$ as a function of the time step $t$. $D^2$ is scaled by $D^2$ at the time step $t$ = 1 for data sorting.
The symbol $+$ represents the $D^2$ for isolated cells, which is averaged over 64 cells. The symbol $\times$
represents the $D^2$ for $J_p$ = 2.0.  The symbol $\times\hspace{-1.75ex}+$  represents the $D^2$ for $J_p$ = 0.0. 
(b) $v$ and $P$ as a function of $J_p$.\label{fig:MSD}}
\end{figure}

$D^2$ is shown as a function of $t$ in Fig.~\ref{fig:MSD}(a). $D^2$ for cells with polarized adhesion ($J_p$ = 2.0)
behaves as $D^2$ $\sim$ $t^2$. 
In contrast, $D^2$ for isolated cells and that for cells with isotropic adhesion ($J_p$ = 0.0)
behave as $D^2$ $\sim$ $t$. These contrasting results suggest that  
the model cells collectively acquire propulsion with persistence by using polarized cell-cell adhesion.

The time period of $D^2$ $\sim$ $t^2$ in Fig.~\ref{fig:MSD}(a) is unexpectedly long. This implies 
a stable order in collective motion. 
To directly confirm this ordering due to polarized cell-cell adhesion, 
the average velocity is calculated as
\begin{eqnarray}
v = \frac{1}{NT} \left|\sum_{m,t} \frac{d {\bm R}_m(t)}{dt}\right|,
\end{eqnarray}   
and is plotted as a function of $J_p$ in Fig.~\ref{fig:MSD}(b) with $P$.
$v$ at $J_p$ = 0 is equal to 0. As $J_p$ increases up to 1, $v$ is almost 0.
As $J_p$ further increases beyond $J_p$ = 1, $v$ gradually increases along with $J_p$ and $P$.
These observations indicate that a stable order in 
collective motion occurs for $J_p$ $>$ 1. 
Overall, these results imply that polarized cell-cell adhesion
enables the model cells to switch their motion from random to collective 
at a threshold of $J_p$.

In conclusion, these results provide a theoretical demonstration that polarized cell-cell adhesion 
can function as the source of collective propulsion with persistence. 
This suggests that cells can mutually lead themselves into a state of collective cell migration
using polarized cell-cell adhesion.

\begin{figure}[t] 
\begin{center} 
\includegraphics[width=1.0\linewidth]{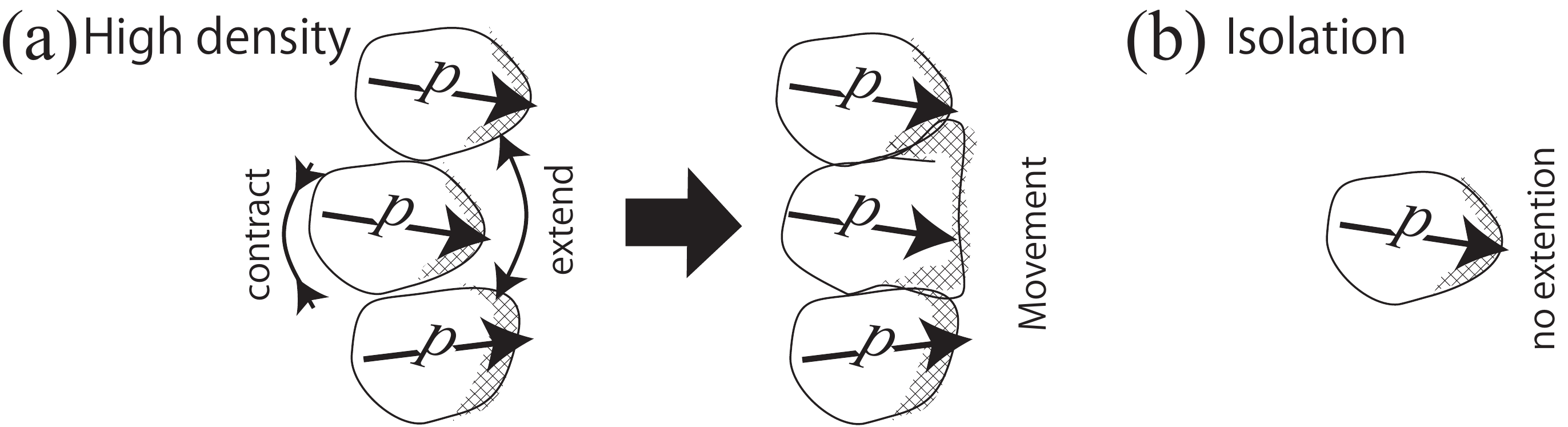} 
\end{center}
\caption{
Schematic diagram of collective propulsion in the cases of
(a) high cell density and (b) an isolated cell. The arrows of ${\bm p}$ 
represent the direction of polarized cell-cell adhesion, and the shaded region 
represents the high-strength region of cell-cell adhesion. 
\label{fig:Mechanism}}
\end{figure}

The emergence of collective propulsion is a notable 
physical phenomenon, but its mechanism of origin is still largely
a mystery. A key to solving this mystery is consideration of the role played by the tension gradient inducing 
propulsion \cite{Levan:1981}. To intuitively approach this question, 
let us consider the periphery tension of a cell (here, we will choose the $m$th cell) 
that is completely surrounded by other cells. For such a cell, the tension term proportional to 
$\sum_{\bm r}{\bm p}_m\cdot{\bm e}_m({\bm r})$ in 
Eq.~\eqref{eq:renormalized-coupling} indicates that the 
tension on the front side is smaller than that on the rear side of the cell. Therefore, 
a cell extends or protrudes from its front side as shown in Fig.~\ref{fig:Mechanism}(a). In 
contrast, a cell comparatively contracts on the rear side.
These extensions and contractions induce the propulsion of a cell in the direction 
of ${\bm p}_m$. This phenomenon is the origin of the emergent propulsion. 
In contrast to this case, since a cell in isolation 
only experiences isotropic tension, as shown in Fig.~\ref{fig:Mechanism}(b), it exhibits a simple random walk.

Since this mechanism only accounts for the emergence of collective propulsion at high density, it is insufficient to explain the ordering of movement shown in Fig.~\ref{fig:MSD}(b). 
The positive feedback control in Fig.~\ref{fig:model}(d) plays a significant role as the origin of the persistence of propulsion to induce this ordering. This can be reasoned as follows. The polarity of adhesion 
$\bm p$ effectively acts as the cell polarity of movement \cite{Kabla:2012} by inducing energy
that is proportional to $\sum_{\bm r}{\bm p}_m\cdot{\bm e}_m({\bm r})$, as discussed above. 
Further, the positive feedback control in Eq.~\eqref{eq:eq_polarity} induces the persistence 
of cell polarity \cite{Czirok:2013}, which is well known 
to induce the ordering of movement \cite{Deseigne:2010}, to ultimately result in the observed ordering.

This emergent collective propulsion may have an essential function in driving collective motion. As described above, cells can theoretically switch their motion from random
to ordering by utilizing cell-cell adhesion. Indeed, polarization in cell-cell adhesion has been shown to arise 
in the aggregating process of {\it Dictyostelium discoideum} \cite{Coates:2001}, which might have functioned as a trigger
of collective motion in evolutionary history. Since confirmation of the function of polarization throughout evolutionary history is difficult, further theoretical exploration of these relationships controlling for the effects of chemotaxis would be an important topic of future research.



This work is supported by JSPS KAKENHI Grant Number 15K17740.
The author would like to thank 
Ryosuke Ishiwata, Hidekazu Kuwayama, Daisuke Mashiko, Shunsuke Yabunaka, and Kenichi Hironaka
for meaningful discussions. 
The author also thanks Koichi Fujimoto, Macoto Kikuchi, and Hajime Yoshino for their generous support.

\begin{thebibliography}{42}%
\makeatletter
\providecommand \@ifxundefined [1]{%
 \@ifx{#1\undefined}
}%
\providecommand \@ifnum [1]{%
 \ifnum #1\expandafter \@firstoftwo
 \else \expandafter \@secondoftwo
 \fi
}%
\providecommand \@ifx [1]{%
 \ifx #1\expandafter \@firstoftwo
 \else \expandafter \@secondoftwo
 \fi
}%
\providecommand \natexlab [1]{#1}%
\providecommand \enquote  [1]{``#1''}%
\providecommand \bibnamefont  [1]{#1}%
\providecommand \bibfnamefont [1]{#1}%
\providecommand \citenamefont [1]{#1}%
\providecommand \href@noop [0]{\@secondoftwo}%
\providecommand \href [0]{\begingroup \@sanitize@url \@href}%
\providecommand \@href[1]{\@@startlink{#1}\@@href}%
\providecommand \@@href[1]{\endgroup#1\@@endlink}%
\providecommand \@sanitize@url [0]{\catcode `\\12\catcode `\$12\catcode
  `\&12\catcode `\#12\catcode `\^12\catcode `\_12\catcode `\%12\relax}%
\providecommand \@@startlink[1]{}%
\providecommand \@@endlink[0]{}%
\providecommand \url  [0]{\begingroup\@sanitize@url \@url }%
\providecommand \@url [1]{\endgroup\@href {#1}{\urlprefix }}%
\providecommand \urlprefix  [0]{URL }%
\providecommand \Eprint [0]{\href }%
\providecommand \doibase [0]{http://dx.doi.org/}%
\providecommand \selectlanguage [0]{\@gobble}%
\providecommand \bibinfo  [0]{\@secondoftwo}%
\providecommand \bibfield  [0]{\@secondoftwo}%
\providecommand \translation [1]{[#1]}%
\providecommand \BibitemOpen [0]{}%
\providecommand \bibitemStop [0]{}%
\providecommand \bibitemNoStop [0]{.\EOS\space}%
\providecommand \EOS [0]{\spacefactor3000\relax}%
\providecommand \BibitemShut  [1]{\csname bibitem#1\endcsname}%
\let\auto@bib@innerbib\@empty
\bibitem [{\citenamefont {Weijer}(2015)}]{Weijer:2009}%
  \BibitemOpen
  \bibfield  {author} {\bibinfo {author} {\bibfnamefont {C.~J.}\ \bibnamefont
  {Weijer}},\ }\href@noop {} {\bibfield  {journal} {\bibinfo  {journal} {J.
  Cell Sci.}\ }\textbf {\bibinfo {volume} {122}},\ \bibinfo {pages} {3215}
  (\bibinfo {year} {2015})}\BibitemShut {NoStop}%
\bibitem [{\citenamefont {Friedl}\ and\ \citenamefont
  {Gilmour}(2009)}]{Friedl:2009}%
  \BibitemOpen
  \bibfield  {author} {\bibinfo {author} {\bibfnamefont {P.}~\bibnamefont
  {Friedl}}\ and\ \bibinfo {author} {\bibfnamefont {D.}~\bibnamefont
  {Gilmour}},\ }\href@noop {} {\bibfield  {journal} {\bibinfo  {journal} {Nat.
  Rev. Mol. Cell Biol.}\ }\textbf {\bibinfo {volume} {10}},\ \bibinfo {pages}
  {445} (\bibinfo {year} {2009})}\BibitemShut {NoStop}%
\bibitem [{\citenamefont {R\mbox{\o}rth}(2009)}]{Rorth:2009}%
  \BibitemOpen
  \bibfield  {author} {\bibinfo {author} {\bibfnamefont {P.}~\bibnamefont
  {R\mbox{\o}rth}},\ }\href@noop {} {\bibfield  {journal} {\bibinfo  {journal}
  {Annu. Rev. Cell Dev. Biol.}\ }\textbf {\bibinfo {volume} {25}},\ \bibinfo
  {pages} {407} (\bibinfo {year} {2009})}\BibitemShut {NoStop}%
\bibitem [{\citenamefont {Haeger}\ \emph {et~al.}(2015)\citenamefont {Haeger},
  \citenamefont {Wolf}, \citenamefont {Zegers},\ and\ \citenamefont
  {Friedl}}]{Haeger:2015}%
  \BibitemOpen
  \bibfield  {author} {\bibinfo {author} {\bibfnamefont {A.}~\bibnamefont
  {Haeger}}, \bibinfo {author} {\bibfnamefont {K.}~\bibnamefont {Wolf}},
  \bibinfo {author} {\bibfnamefont {M.~M.}\ \bibnamefont {Zegers}}, \ and\
  \bibinfo {author} {\bibfnamefont {P.}~\bibnamefont {Friedl}},\ }\href@noop {}
  {\bibfield  {journal} {\bibinfo  {journal} {Trends Cell Biol.}\ }\textbf
  {\bibinfo {volume} {25}},\ \bibinfo {pages} {556} (\bibinfo {year}
  {2015})}\BibitemShut {NoStop}%
\bibitem [{\citenamefont {Sherratt}\ and\ \citenamefont
  {Murray}(1990)}]{Sherratt:1990}%
  \BibitemOpen
  \bibfield  {author} {\bibinfo {author} {\bibfnamefont {J.~A.}\ \bibnamefont
  {Sherratt}}\ and\ \bibinfo {author} {\bibfnamefont {J.~D.}\ \bibnamefont
  {Murray}},\ }\href@noop {} {\bibfield  {journal} {\bibinfo  {journal} {Proc.
  R. Soc. London, Ser. B}\ }\textbf {\bibinfo {volume} {241}},\ \bibinfo
  {pages} {29} (\bibinfo {year} {1990})}\BibitemShut {NoStop}%
\bibitem [{\citenamefont {Szab\'{o}}\ \emph {et~al.}(2006)\citenamefont
  {Szab\'{o}}, \citenamefont {Szollosi}, \citenamefont {Gonci}, \citenamefont
  {Juranyi}, \citenamefont {Selmeczi},\ and\ \citenamefont
  {Vicsek}}]{Szabo:2006}%
  \BibitemOpen
  \bibfield  {author} {\bibinfo {author} {\bibfnamefont {B.}~\bibnamefont
  {Szab\'{o}}}, \bibinfo {author} {\bibfnamefont {G.~J.}\ \bibnamefont
  {Szollosi}}, \bibinfo {author} {\bibfnamefont {B.}~\bibnamefont {Gonci}},
  \bibinfo {author} {\bibfnamefont {Z.}~\bibnamefont {Juranyi}}, \bibinfo
  {author} {\bibfnamefont {D.}~\bibnamefont {Selmeczi}}, \ and\ \bibinfo
  {author} {\bibfnamefont {T.}~\bibnamefont {Vicsek}},\ }\href@noop {}
  {\bibfield  {journal} {\bibinfo  {journal} {Phys.~Rev.~E}\ }\textbf {\bibinfo
  {volume} {74}},\ \bibinfo {pages} {061908} (\bibinfo {year}
  {2006})}\BibitemShut {NoStop}%
\bibitem [{\citenamefont {Lee}\ and\ \citenamefont
  {Wolgemuth}(2011{\natexlab{a}})}]{Lee:2011a}%
  \BibitemOpen
  \bibfield  {author} {\bibinfo {author} {\bibfnamefont {P.}~\bibnamefont
  {Lee}}\ and\ \bibinfo {author} {\bibfnamefont {C.~W.}\ \bibnamefont
  {Wolgemuth}},\ }\href@noop {} {\bibfield  {journal} {\bibinfo  {journal}
  {PloS Comput. Biol.}\ }\textbf {\bibinfo {volume} {7}},\ \bibinfo {pages}
  {e1002007} (\bibinfo {year} {2011}{\natexlab{a}})}\BibitemShut {NoStop}%
\bibitem [{\citenamefont {Lee}\ and\ \citenamefont
  {Wolgemuth}(2011{\natexlab{b}})}]{Lee:2011b}%
  \BibitemOpen
  \bibfield  {author} {\bibinfo {author} {\bibfnamefont {P.}~\bibnamefont
  {Lee}}\ and\ \bibinfo {author} {\bibfnamefont {C.}~\bibnamefont
  {Wolgemuth}},\ }\href@noop {} {\bibfield  {journal} {\bibinfo  {journal}
  {Phys. Rev. E}\ }\textbf {\bibinfo {volume} {83}},\ \bibinfo {pages} {061920}
  (\bibinfo {year} {2011}{\natexlab{b}})}\BibitemShut {NoStop}%
\bibitem [{\citenamefont {Vicsek}\ and\ \citenamefont
  {Zafeiris}(2012)}]{Vicsek:2012}%
  \BibitemOpen
  \bibfield  {author} {\bibinfo {author} {\bibfnamefont {T.}~\bibnamefont
  {Vicsek}}\ and\ \bibinfo {author} {\bibfnamefont {A.}~\bibnamefont
  {Zafeiris}},\ }\href@noop {} {\bibfield  {journal} {\bibinfo  {journal}
  {Phys. Rep.}\ }\textbf {\bibinfo {volume} {517}},\ \bibinfo {pages} {71}
  (\bibinfo {year} {2012})}\BibitemShut {NoStop}%
\bibitem [{\citenamefont {Basan}\ \emph {et~al.}(2013)\citenamefont {Basan},
  \citenamefont {Elgeti}, \citenamefont {Hannezo}, \citenamefont {Rappel},\
  and\ \citenamefont {Levine}}]{Basan:2013}%
  \BibitemOpen
  \bibfield  {author} {\bibinfo {author} {\bibfnamefont {M.}~\bibnamefont
  {Basan}}, \bibinfo {author} {\bibfnamefont {J.}~\bibnamefont {Elgeti}},
  \bibinfo {author} {\bibfnamefont {E.}~\bibnamefont {Hannezo}}, \bibinfo
  {author} {\bibfnamefont {W.-J.}\ \bibnamefont {Rappel}}, \ and\ \bibinfo
  {author} {\bibfnamefont {H.}~\bibnamefont {Levine}},\ }\href@noop {}
  {\bibfield  {journal} {\bibinfo  {journal} {Proc. Natl. Acad. Sci. USA}\
  }\textbf {\bibinfo {volume} {110}},\ \bibinfo {pages} {2452} (\bibinfo {year}
  {2013})}\BibitemShut {NoStop}%
\bibitem [{\citenamefont {Marchetti}\ \emph {et~al.}(2013)\citenamefont
  {Marchetti}, \citenamefont {Joanny}, \citenamefont {Ramaswamy}, \citenamefont
  {Liverpool}, \citenamefont {Prost}, \citenamefont {Rao},\ and\ \citenamefont
  {Simha}}]{Marchetti:2013}%
  \BibitemOpen
  \bibfield  {author} {\bibinfo {author} {\bibfnamefont {M.~C.}\ \bibnamefont
  {Marchetti}}, \bibinfo {author} {\bibfnamefont {J.~F.}\ \bibnamefont
  {Joanny}}, \bibinfo {author} {\bibfnamefont {S.}~\bibnamefont {Ramaswamy}},
  \bibinfo {author} {\bibfnamefont {T.~B.}\ \bibnamefont {Liverpool}}, \bibinfo
  {author} {\bibfnamefont {J.}~\bibnamefont {Prost}}, \bibinfo {author}
  {\bibfnamefont {M.}~\bibnamefont {Rao}}, \ and\ \bibinfo {author}
  {\bibfnamefont {R.~A.}\ \bibnamefont {Simha}},\ }\href@noop {} {\bibfield
  {journal} {\bibinfo  {journal} {Rev. Mod. Phys.}\ }\textbf {\bibinfo {volume}
  {85}},\ \bibinfo {pages} {1143} (\bibinfo {year} {2013})}\BibitemShut
  {NoStop}%
\bibitem [{\citenamefont {Sep\'{u}lveda}\ \emph {et~al.}(2013)\citenamefont
  {Sep\'{u}lveda}, \citenamefont {Petitjean}, \citenamefont {Cochet},
  \citenamefont {Grasland-Mongrain}, \citenamefont {Silberzan},\ and\
  \citenamefont {Hakim}}]{Sepulveda:2013}%
  \BibitemOpen
  \bibfield  {author} {\bibinfo {author} {\bibfnamefont {N.}~\bibnamefont
  {Sep\'{u}lveda}}, \bibinfo {author} {\bibfnamefont {L.}~\bibnamefont
  {Petitjean}}, \bibinfo {author} {\bibfnamefont {O.}~\bibnamefont {Cochet}},
  \bibinfo {author} {\bibfnamefont {E.}~\bibnamefont {Grasland-Mongrain}},
  \bibinfo {author} {\bibfnamefont {P.}~\bibnamefont {Silberzan}}, \ and\
  \bibinfo {author} {\bibfnamefont {V.}~\bibnamefont {Hakim}},\ }\href@noop {}
  {\bibfield  {journal} {\bibinfo  {journal} {PloS Comp. Biol.}\ }\textbf
  {\bibinfo {volume} {9}},\ \bibinfo {pages} {e1002944} (\bibinfo {year}
  {2013})}\BibitemShut {NoStop}%
\bibitem [{\citenamefont {Li}\ and\ \citenamefont {Sun}(2014)}]{Li:2014}%
  \BibitemOpen
  \bibfield  {author} {\bibinfo {author} {\bibfnamefont {B.}~\bibnamefont
  {Li}}\ and\ \bibinfo {author} {\bibfnamefont {S.~X.}\ \bibnamefont {Sun}},\
  }\href@noop {} {\bibfield  {journal} {\bibinfo  {journal} {Biophys. J.}\
  }\textbf {\bibinfo {volume} {107}},\ \bibinfo {pages} {1532} (\bibinfo {year}
  {2014})}\BibitemShut {NoStop}%
\bibitem [{\citenamefont {Londono}\ \emph {et~al.}(2014)\citenamefont
  {Londono}, \citenamefont {Loureiro}, \citenamefont {Slater}, \citenamefont
  {L\"{u}cker}, \citenamefont {Soleasa}, \citenamefont {Sathananthan},
  \citenamefont {Aitchison}, \citenamefont {Kabla},\ and\ \citenamefont
  {McGuigan}}]{Londono:2014}%
  \BibitemOpen
  \bibfield  {author} {\bibinfo {author} {\bibfnamefont {C.}~\bibnamefont
  {Londono}}, \bibinfo {author} {\bibfnamefont {M.~J.}\ \bibnamefont
  {Loureiro}}, \bibinfo {author} {\bibfnamefont {B.}~\bibnamefont {Slater}},
  \bibinfo {author} {\bibfnamefont {P.~B.}\ \bibnamefont {L\"{u}cker}},
  \bibinfo {author} {\bibfnamefont {J.}~\bibnamefont {Soleasa}}, \bibinfo
  {author} {\bibfnamefont {S.}~\bibnamefont {Sathananthan}}, \bibinfo {author}
  {\bibfnamefont {J.~S.}\ \bibnamefont {Aitchison}}, \bibinfo {author}
  {\bibfnamefont {A.~J.}\ \bibnamefont {Kabla}}, \ and\ \bibinfo {author}
  {\bibfnamefont {A.~P.}\ \bibnamefont {McGuigan}},\ }\href@noop {} {\bibfield
  {journal} {\bibinfo  {journal} {Proc. Natl. Acad. Sci. USA}\ }\textbf
  {\bibinfo {volume} {111}},\ \bibinfo {pages} {1807} (\bibinfo {year}
  {2014})}\BibitemShut {NoStop}%
\bibitem [{\citenamefont {Camley}\ \emph {et~al.}(2016)\citenamefont {Camley},
  \citenamefont {Zimmermann}, \citenamefont {Levine},\ and\ \citenamefont
  {Rappel}}]{Camley:2016}%
  \BibitemOpen
  \bibfield  {author} {\bibinfo {author} {\bibfnamefont {B.~A.}\ \bibnamefont
  {Camley}}, \bibinfo {author} {\bibfnamefont {J.}~\bibnamefont {Zimmermann}},
  \bibinfo {author} {\bibfnamefont {H.}~\bibnamefont {Levine}}, \ and\ \bibinfo
  {author} {\bibfnamefont {W.-J.}\ \bibnamefont {Rappel}},\ }\href@noop {}
  {\bibfield  {journal} {\bibinfo  {journal} {Phys. Rev. Lett.}\ }\textbf
  {\bibinfo {volume} {116}},\ \bibinfo {pages} {098101} (\bibinfo {year}
  {2016})}\BibitemShut {NoStop}%
\bibitem [{\citenamefont {Kabla}(2012)}]{Kabla:2012}%
  \BibitemOpen
  \bibfield  {author} {\bibinfo {author} {\bibfnamefont {A.~J.}\ \bibnamefont
  {Kabla}},\ }\href@noop {} {\bibfield  {journal} {\bibinfo  {journal} {J. R.
  Soc. Interface}\ }\textbf {\bibinfo {volume} {9}},\ \bibinfo {pages} {3268}
  (\bibinfo {year} {2012})}\BibitemShut {NoStop}%
\bibitem [{\citenamefont {Lo}\ \emph {et~al.}(2000)\citenamefont {Lo},
  \citenamefont {Wang}, \citenamefont {Dembo},\ and\ \citenamefont
  {li~Wang}}]{Lo:2000}%
  \BibitemOpen
  \bibfield  {author} {\bibinfo {author} {\bibfnamefont {C.-M.}\ \bibnamefont
  {Lo}}, \bibinfo {author} {\bibfnamefont {H.-B.}\ \bibnamefont {Wang}},
  \bibinfo {author} {\bibfnamefont {M.}~\bibnamefont {Dembo}}, \ and\ \bibinfo
  {author} {\bibfnamefont {Y.}~\bibnamefont {li~Wang}},\ }\href@noop {}
  {\bibfield  {journal} {\bibinfo  {journal} {Biophys. J.}\ }\textbf {\bibinfo
  {volume} {79}},\ \bibinfo {pages} {144} (\bibinfo {year} {2000})}\BibitemShut
  {NoStop}%
\bibitem [{\citenamefont {Carter}(1967)}]{Carter:1967}%
  \BibitemOpen
  \bibfield  {author} {\bibinfo {author} {\bibfnamefont {S.~B.}\ \bibnamefont
  {Carter}},\ }\href@noop {} {\bibfield  {journal} {\bibinfo  {journal}
  {Nature}\ }\textbf {\bibinfo {volume} {5073}},\ \bibinfo {pages} {256}
  (\bibinfo {year} {1967})}\BibitemShut {NoStop}%
\bibitem [{\citenamefont {M\'{e}hes}\ and\ \citenamefont
  {Vicsek}(2013)}]{Mehes:2013}%
  \BibitemOpen
  \bibfield  {author} {\bibinfo {author} {\bibfnamefont {E.}~\bibnamefont
  {M\'{e}hes}}\ and\ \bibinfo {author} {\bibfnamefont {T.}~\bibnamefont
  {Vicsek}},\ }\href@noop {} {\bibfield  {journal} {\bibinfo  {journal}
  {Comput.. Adapt. Syst. Mod.}\ }\textbf {\bibinfo {volume} {1}},\ \bibinfo
  {pages} {4} (\bibinfo {year} {2013})}\BibitemShut {NoStop}%
\bibitem [{\citenamefont {Pfeffer}(1884)}]{Pfeffer:1884}%
  \BibitemOpen
  \bibfield  {author} {\bibinfo {author} {\bibfnamefont {W.}~\bibnamefont
  {Pfeffer}},\ }\href@noop {} {\bibfield  {journal} {\bibinfo  {journal}
  {Umtersuch. Bot. Inst. T\mbox{\"u}bingen.}\ }\textbf {\bibinfo {volume}
  {1}},\ \bibinfo {pages} {363} (\bibinfo {year} {1884})}\BibitemShut {NoStop}%
\bibitem [{\citenamefont {Bonner}(2009)}]{Bonner:2009}%
  \BibitemOpen
  \bibfield  {author} {\bibinfo {author} {\bibfnamefont {J.~T.}\ \bibnamefont
  {Bonner}},\ }\href@noop {} {\emph {\bibinfo {title} {The Social Amoebae: The
  Biology of Cellular Slime Molds}}}\ (\bibinfo  {publisher} {Princeton
  University Press, Princeton},\ \bibinfo {year} {2009})\BibitemShut {NoStop}%
\bibitem [{\citenamefont {Carmona-Fontaine}\ \emph {et~al.}(2008)\citenamefont
  {Carmona-Fontaine}, \citenamefont {Matthews}, \citenamefont {Kuriyama},
  \citenamefont {Moreno}, \citenamefont {Dunn}, \citenamefont {Parsons},
  \citenamefont {Stern},\ and\ \citenamefont {Mayor}}]{Carmona-Fontaine:2008}%
  \BibitemOpen
  \bibfield  {author} {\bibinfo {author} {\bibfnamefont {C.}~\bibnamefont
  {Carmona-Fontaine}}, \bibinfo {author} {\bibfnamefont {H.~K.}\ \bibnamefont
  {Matthews}}, \bibinfo {author} {\bibfnamefont {S.}~\bibnamefont {Kuriyama}},
  \bibinfo {author} {\bibfnamefont {M.}~\bibnamefont {Moreno}}, \bibinfo
  {author} {\bibfnamefont {G.~A.}\ \bibnamefont {Dunn}}, \bibinfo {author}
  {\bibfnamefont {M.}~\bibnamefont {Parsons}}, \bibinfo {author} {\bibfnamefont
  {C.~D.}\ \bibnamefont {Stern}}, \ and\ \bibinfo {author} {\bibfnamefont
  {R.}~\bibnamefont {Mayor}},\ }\href@noop {} {\bibfield  {journal} {\bibinfo
  {journal} {Nature (London)}\ }\textbf {\bibinfo {volume} {456}},\ \bibinfo
  {pages} {957} (\bibinfo {year} {2008})}\BibitemShut {NoStop}%
\bibitem [{\citenamefont {Swaney}\ \emph {et~al.}(2010)\citenamefont {Swaney},
  \citenamefont {Huang},\ and\ \citenamefont {Devreotes}}]{Swaney:2010}%
  \BibitemOpen
  \bibfield  {author} {\bibinfo {author} {\bibfnamefont {K.~F.}\ \bibnamefont
  {Swaney}}, \bibinfo {author} {\bibfnamefont {C.-H.}\ \bibnamefont {Huang}}, \
  and\ \bibinfo {author} {\bibfnamefont {P.~N.}\ \bibnamefont {Devreotes}},\
  }\href@noop {} {\bibfield  {journal} {\bibinfo  {journal} {Annu. Rev.
  Biophys.}\ }\textbf {\bibinfo {volume} {39}},\ \bibinfo {pages} {265}
  (\bibinfo {year} {2010})}\BibitemShut {NoStop}%
\bibitem [{\citenamefont {Takeichi}(2014)}]{Takeichi:2014}%
  \BibitemOpen
  \bibfield  {author} {\bibinfo {author} {\bibfnamefont {M.}~\bibnamefont
  {Takeichi}},\ }\href@noop {} {\bibfield  {journal} {\bibinfo  {journal} {Nat.
  Rev. Mol. Cell. Biol.}\ }\textbf {\bibinfo {volume} {15}},\ \bibinfo {pages}
  {397} (\bibinfo {year} {2014})}\BibitemShut {NoStop}%
\bibitem [{\citenamefont {Dumortier}\ \emph {et~al.}(2012)\citenamefont
  {Dumortier}, \citenamefont {Martin}, \citenamefont {Meyer}, \citenamefont
  {Rosa},\ and\ \citenamefont {David}}]{Dumortier:2012}%
  \BibitemOpen
  \bibfield  {author} {\bibinfo {author} {\bibfnamefont {J.~G.}\ \bibnamefont
  {Dumortier}}, \bibinfo {author} {\bibfnamefont {S.}~\bibnamefont {Martin}},
  \bibinfo {author} {\bibfnamefont {D.}~\bibnamefont {Meyer}}, \bibinfo
  {author} {\bibfnamefont {F.~M.}\ \bibnamefont {Rosa}}, \ and\ \bibinfo
  {author} {\bibfnamefont {N.~B.}\ \bibnamefont {David}},\ }\href@noop {}
  {\bibfield  {journal} {\bibinfo  {journal} {ProcProc}\ }\textbf {\bibinfo
  {volume} {109}},\ \bibinfo {pages} {16945} (\bibinfo {year}
  {2012})}\BibitemShut {NoStop}%
\bibitem [{\citenamefont {Cai}\ \emph {et~al.}(2014)\citenamefont {Cai},
  \citenamefont {Chen}, \citenamefont {Prasad}, \citenamefont {He},
  \citenamefont {Wang}, \citenamefont {Choesmel-Cadamuro}, \citenamefont
  {Sawyer}, \citenamefont {Danuser},\ and\ \citenamefont {Montell}}]{Cai:2014}%
  \BibitemOpen
  \bibfield  {author} {\bibinfo {author} {\bibfnamefont {D.}~\bibnamefont
  {Cai}}, \bibinfo {author} {\bibfnamefont {S.-C.}\ \bibnamefont {Chen}},
  \bibinfo {author} {\bibfnamefont {M.}~\bibnamefont {Prasad}}, \bibinfo
  {author} {\bibfnamefont {L.}~\bibnamefont {He}}, \bibinfo {author}
  {\bibfnamefont {X.}~\bibnamefont {Wang}}, \bibinfo {author} {\bibfnamefont
  {V.}~\bibnamefont {Choesmel-Cadamuro}}, \bibinfo {author} {\bibfnamefont
  {J.~K.}\ \bibnamefont {Sawyer}}, \bibinfo {author} {\bibfnamefont
  {G.}~\bibnamefont {Danuser}}, \ and\ \bibinfo {author} {\bibfnamefont
  {D.~J.}\ \bibnamefont {Montell}},\ }\href@noop {} {\bibfield  {journal}
  {\bibinfo  {journal} {Cell}\ }\textbf {\bibinfo {volume} {157}},\ \bibinfo
  {pages} {1146} (\bibinfo {year} {2014})}\BibitemShut {NoStop}%
\bibitem [{\citenamefont {Friedl}\ and\ \citenamefont
  {Wolf}(2003)}]{Friedl:2003}%
  \BibitemOpen
  \bibfield  {author} {\bibinfo {author} {\bibfnamefont {P.}~\bibnamefont
  {Friedl}}\ and\ \bibinfo {author} {\bibfnamefont {K.}~\bibnamefont {Wolf}},\
  }\href@noop {} {\bibfield  {journal} {\bibinfo  {journal} {Nat. Rev. Cancer}\
  }\textbf {\bibinfo {volume} {3}},\ \bibinfo {pages} {362} (\bibinfo {year}
  {2003})}\BibitemShut {NoStop}%
\bibitem [{\citenamefont {Beug}\ \emph {et~al.}(1973)\citenamefont {Beug},
  \citenamefont {Katz},\ and\ \citenamefont {Gerisch}}]{Beug:1973}%
  \BibitemOpen
  \bibfield  {author} {\bibinfo {author} {\bibfnamefont {H.}~\bibnamefont
  {Beug}}, \bibinfo {author} {\bibfnamefont {F.~E.}\ \bibnamefont {Katz}}, \
  and\ \bibinfo {author} {\bibfnamefont {G.}~\bibnamefont {Gerisch}},\
  }\href@noop {} {\bibfield  {journal} {\bibinfo  {journal} {J. Cell Biol.}\
  }\textbf {\bibinfo {volume} {56}},\ \bibinfo {pages} {647} (\bibinfo {year}
  {1973})}\BibitemShut {NoStop}%
\bibitem [{\citenamefont {M\"{u}ller}\ and\ \citenamefont
  {Gerisch}(1978)}]{Muller:1978}%
  \BibitemOpen
  \bibfield  {author} {\bibinfo {author} {\bibfnamefont {K.}~\bibnamefont
  {M\"{u}ller}}\ and\ \bibinfo {author} {\bibfnamefont {G.}~\bibnamefont
  {Gerisch}},\ }\href@noop {} {\bibfield  {journal} {\bibinfo  {journal}
  {Nature (London)}\ }\textbf {\bibinfo {volume} {274}},\ \bibinfo {pages}
  {445} (\bibinfo {year} {1978})}\BibitemShut {NoStop}%
\bibitem [{\citenamefont {Theveneau}(2010)}]{Theveneau:2010}%
  \BibitemOpen
  \bibfield  {author} {\bibinfo {author} {\bibfnamefont {E.}~\bibnamefont
  {Theveneau}},\ }\href@noop {} {\bibfield  {journal} {\bibinfo  {journal}
  {Dev. Cell.}\ }\textbf {\bibinfo {volume} {19}},\ \bibinfo {pages} {39}
  (\bibinfo {year} {2010})}\BibitemShut {NoStop}%
\bibitem [{\citenamefont {Sesaki}\ and\ \citenamefont
  {Siu}(1996)}]{Sesaki:1996}%
  \BibitemOpen
  \bibfield  {author} {\bibinfo {author} {\bibfnamefont {H.}~\bibnamefont
  {Sesaki}}\ and\ \bibinfo {author} {\bibfnamefont {C.-H.}\ \bibnamefont
  {Siu}},\ }\href@noop {} {\bibfield  {journal} {\bibinfo  {journal} {Develop.
  Biol.}\ }\textbf {\bibinfo {volume} {177}},\ \bibinfo {pages} {504} (\bibinfo
  {year} {1996})}\BibitemShut {NoStop}%
\bibitem [{\citenamefont {Graner}\ and\ \citenamefont
  {Glazier}(1992)}]{Graner:1992}%
  \BibitemOpen
  \bibfield  {author} {\bibinfo {author} {\bibfnamefont {F.}~\bibnamefont
  {Graner}}\ and\ \bibinfo {author} {\bibfnamefont {J.~A.}\ \bibnamefont
  {Glazier}},\ }\href@noop {} {\bibfield  {journal} {\bibinfo  {journal}
  {Phys.~Rev.~Lett.}\ }\textbf {\bibinfo {volume} {69}},\ \bibinfo {pages}
  {2013} (\bibinfo {year} {1992})}\BibitemShut {NoStop}%
\bibitem [{\citenamefont {Anderson}\ \emph {et~al.}(2007)\citenamefont
  {Anderson}, \citenamefont {Chaplain},\ and\ \citenamefont
  {Rejniak}}]{Anderson:2007}%
  \BibitemOpen
  \bibfield  {author} {\bibinfo {author} {\bibfnamefont {A.~R.~A.}\
  \bibnamefont {Anderson}}, \bibinfo {author} {\bibfnamefont {M.~A.~J.}\
  \bibnamefont {Chaplain}}, \ and\ \bibinfo {author} {\bibfnamefont {K.~A.}\
  \bibnamefont {Rejniak}},\ }\href@noop {} {\emph {\bibinfo {title}
  {Single-Cell-Based Models in Biology and Medicine}}}\ (\bibinfo  {publisher}
  {Birkhauser Verlag AG, Basel},\ \bibinfo {year} {2007})\BibitemShut {NoStop}%
\bibitem [{\citenamefont {Glazier}\ and\ \citenamefont
  {Graner}(1993)}]{Glazier:1993}%
  \BibitemOpen
  \bibfield  {author} {\bibinfo {author} {\bibfnamefont {J.~A.}\ \bibnamefont
  {Glazier}}\ and\ \bibinfo {author} {\bibfnamefont {F.}~\bibnamefont
  {Graner}},\ }\href@noop {} {\bibfield  {journal} {\bibinfo  {journal} {Phys.
  Rev. E}\ }\textbf {\bibinfo {volume} {47}},\ \bibinfo {pages} {2128}
  (\bibinfo {year} {1993})}\BibitemShut {NoStop}%
\bibitem [{\citenamefont {Graner}(1993)}]{Graner:1993}%
  \BibitemOpen
  \bibfield  {author} {\bibinfo {author} {\bibfnamefont {F.}~\bibnamefont
  {Graner}},\ }\href@noop {} {\bibfield  {journal} {\bibinfo  {journal} {J.
  Theor. Biol.}\ }\textbf {\bibinfo {volume} {164}},\ \bibinfo {pages} {455}
  (\bibinfo {year} {1993})}\BibitemShut {NoStop}%
\bibitem [{\citenamefont {Zajac}\ \emph {et~al.}(2002)\citenamefont {Zajac},
  \citenamefont {Jonesa},\ and\ \citenamefont {Glazier}}]{Zajac:2002}%
  \BibitemOpen
  \bibfield  {author} {\bibinfo {author} {\bibfnamefont {M.}~\bibnamefont
  {Zajac}}, \bibinfo {author} {\bibfnamefont {G.~L.}\ \bibnamefont {Jonesa}}, \
  and\ \bibinfo {author} {\bibfnamefont {J.~A.}\ \bibnamefont {Glazier}},\
  }\href@noop {} {\bibfield  {journal} {\bibinfo  {journal} {J. Theor. Biol.}\
  }\textbf {\bibinfo {volume} {222}},\ \bibinfo {pages} {247} (\bibinfo {year}
  {2002})}\BibitemShut {NoStop}%
\bibitem [{\citenamefont {Vroomans}\ \emph {et~al.}(2015)\citenamefont
  {Vroomans}, \citenamefont {Hogeweg},\ and\ \citenamefont {ten
  Tusscher}}]{Vroomans:2015}%
  \BibitemOpen
  \bibfield  {author} {\bibinfo {author} {\bibfnamefont {R.~M.~A.}\
  \bibnamefont {Vroomans}}, \bibinfo {author} {\bibfnamefont {P.}~\bibnamefont
  {Hogeweg}}, \ and\ \bibinfo {author} {\bibfnamefont {K.~H. W.~J.}\
  \bibnamefont {ten Tusscher}},\ }\href@noop {} {\bibfield  {journal} {\bibinfo
   {journal} {PLoS Comput. Biol.}\ }\textbf {\bibinfo {volume} {11}},\ \bibinfo
  {pages} {e1004092} (\bibinfo {year} {2015})}\BibitemShut {NoStop}%
\bibitem [{\citenamefont {Matsushita}(2017)}]{Matsushita:2017}%
  \BibitemOpen
  \bibfield  {author} {\bibinfo {author} {\bibfnamefont {K.}~\bibnamefont
  {Matsushita}},\ }\href@noop {} {\bibfield  {journal} {\bibinfo  {journal}
  {Phys. Rev. E.}\ }\textbf {\bibinfo {volume} {95}},\ \bibinfo {pages}
  {032415} (\bibinfo {year} {2017})}\BibitemShut {NoStop}%
\bibitem [{\citenamefont {Levan}(1981)}]{Levan:1981}%
  \BibitemOpen
  \bibfield  {author} {\bibinfo {author} {\bibfnamefont {M.~D.}\ \bibnamefont
  {Levan}},\ }\href@noop {} {\bibfield  {journal} {\bibinfo  {journal} {J.
  Coll. Int. Sci.}\ }\textbf {\bibinfo {volume} {83}},\ \bibinfo {pages} {11}
  (\bibinfo {year} {1981})}\BibitemShut {NoStop}%
\bibitem [{\citenamefont {Czir\'ok}\ \emph {et~al.}(2013)\citenamefont
  {Czir\'ok}, \citenamefont {Varga}, \citenamefont {M\'ehes},\ and\
  \citenamefont {Szab\'o}}]{Czirok:2013}%
  \BibitemOpen
  \bibfield  {author} {\bibinfo {author} {\bibfnamefont {A.}~\bibnamefont
  {Czir\'ok}}, \bibinfo {author} {\bibfnamefont {K.}~\bibnamefont {Varga}},
  \bibinfo {author} {\bibfnamefont {E.}~\bibnamefont {M\'ehes}}, \ and\
  \bibinfo {author} {\bibfnamefont {A.}~\bibnamefont {Szab\'o}},\ }\href@noop
  {} {\bibfield  {journal} {\bibinfo  {journal} {New J}\ }\textbf {\bibinfo
  {volume} {15}},\ \bibinfo {pages} {075006} (\bibinfo {year}
  {2013})}\BibitemShut {NoStop}%
\bibitem [{\citenamefont {Deseigne}\ \emph {et~al.}(2010)\citenamefont
  {Deseigne}, \citenamefont {Dauchot},\ and\ \citenamefont
  {Chat\'{e}}}]{Deseigne:2010}%
  \BibitemOpen
  \bibfield  {author} {\bibinfo {author} {\bibfnamefont {J.}~\bibnamefont
  {Deseigne}}, \bibinfo {author} {\bibfnamefont {O.}~\bibnamefont {Dauchot}}, \
  and\ \bibinfo {author} {\bibfnamefont {H.}~\bibnamefont {Chat\'{e}}},\
  }\href@noop {} {\bibfield  {journal} {\bibinfo  {journal} {Phys. Rev. Lett.}\
  }\textbf {\bibinfo {volume} {105}},\ \bibinfo {pages} {098001} (\bibinfo
  {year} {2010})}\BibitemShut {NoStop}%
\bibitem [{\citenamefont {Coates}\ and\ \citenamefont
  {Harwood}(2001)}]{Coates:2001}%
  \BibitemOpen
  \bibfield  {author} {\bibinfo {author} {\bibfnamefont {J.~C.}\ \bibnamefont
  {Coates}}\ and\ \bibinfo {author} {\bibfnamefont {A.~J.}\ \bibnamefont
  {Harwood}},\ }\href@noop {} {\bibfield  {journal} {\bibinfo  {journal} {J.
  Cell Sci.}\ }\textbf {\bibinfo {volume} {114}},\ \bibinfo {pages} {4349}
  (\bibinfo {year} {2001})}\BibitemShut {NoStop}%
\end{thebibliography}
%
\end{document}